\documentclass[11pt,a4paper]{article}
\usepackage{graphicx}
\begin{document}
\sloppy

\twocolumn[
\begin{center}
{\large\bf Microvariability of Line Profiles in the Spectra of OB Stars II: 
    $\delta\,$Ori~A}
\end{center}

\begin{center}
{\large\bf A.F.Kholtygin$^{1,*}$, T.E.Burlakova$^{2,3}$, S.N.Fabrika$^{3}$,
     G.G.Valyavin$^{2,3}$, M.V.Yushkin$^{3}$}
\end{center}

\begin{center}
{\it $^1$Sobolev Astronomical Institute, St. Petersburg State University, 
Bibliotechnaya pl. 2, Petrodvorets, 198904 Russia 
$^*$E-Mail: afk@theor1.astro.spbu.ru} \\
{\it $^2$ Bohyunsan Optical Astronomy Observatory, Jaceon P.O.B.,
Young-chun, Kyung-pook, 770-820, Republic of Korea
}
{\it $^3$ Special Astrophysical Observatory, Russian Academy of Sciences,
Nizhni.. Arkhyz, Karacha.. -Cherkessian Republic, 357147 Russia
}
\end{center}

{\small
\noindent
{\bf Abstract}--We have studied variability of the spectral lines of the OB star 
  $\delta\,$Ori~A--the brightest component 
  of the $\delta\,$Ori triple system. Forty spectra with signal-to-noise ratios 
   $\approx$500--800 and a time resolution of four minutes were obtained. 
We detected variability in the HeII$\,\lambda\,4686$, HeI$\,\lambda\,$4713 and 
H$\beta$ absorption and the CIII$\,\lambda\,$5696 emission line profiles. 
The amplitude of the variability is $\approx$(0.5--1)\% of the continuum 
intensity. 
The dynamical wavelet spectrum of the profile variations reveals large-scale 
components in the interval 25--50 km/s that move within the -$V\sin i$ to 
$V\sin i$ band for the primary star of the system, Aa$^1$, with 
a band crossing time of 4$^h$--5$^h$. However, some of the variable features go 
outside the band, presumably due to either imhomogeneities in the stellar wind 
from $\delta\,$Ori Aa$^1$ or non-radial pulsations of the weaker 
components of the system, Aa$^2$ or Ab. The detected variability may by 
cyclic with a period of $\approx 4^h$ . We suggest that it is associated with 
non-radial pulsations of the primary in the sector mode $(l, m)=(2,-2)$. 
}\\
]                             

\section{Introduction}

Spectral observations of hot stars at UV~\cite{Kaper97,Kaper99}, 
optical~\cite{Kaper97,Kaufer96,Kaper99,LM99,deJong99,deJong01}, 
and X-ray~\cite{Kahn01,Miller02} wavelengths indicate 
the presence of structures in the atmospheres of these 
stars with various sizes and densities, and with lifetimes 
from less than an hour to several days. Variations 
of the line profiles in the spectra of OB stars 
are mainly regular or quasi-regular. The spectra of B 
stars and six O stars display regular short-period (3 -- 12 h) profile 
variations for numerous lines (in particular, 
HeI) associated with the non-radial pulsations of 
these stars~\cite{deJong01}. 

Weak magnetic fields (several hundred Gauss at 
the stellar surface) may be one factor favoring the formation 
of large-scale structures in the atmospheres 
of O stars\cite{Donati01}. Currently, magnetic fields have been 
detected with sufficient reliability only in one O star, 
$\theta^{1}\,$Ori~C \cite{Donati02}, and several early 
B subtype stars\cite{Neiner02}. 
Unlike Bstars, the amplitudes of the line profile variations 
in the spectra of O stars are relatively small 
(1\%--3\%; see, for example, [12]), so it it is more 
appropriate to call this line profile {\it microvariability}. 
To detect such variability and to clarify its nature, 
observations with high time and spectral resolution 
and high signal-to-noise ratios ($\ge 300$) are needed. 

Here, we continue our study of the microvariability 
of hot stars begun in [12], when we analyzed line-
profile variability in the spectra of the supergiants  
19 Cep, $\alpha\,$Cam ¨ $\lambda\,$Ori~A (O9.5II). 

We now present the results of spectral observations 
of the spectral triple system $\delta\,$Ori~A, whose 
primary displays the same spectral type as $\lambda\,$Ori~A -- 
O9.5. Section 2 contains general information about 
the $\delta\,$Ori~A system, and Section 3 describes the observations 
and reduction of the spectra. Section 4 
presents the methods used in and the results of our 
analysis of line-profile variability in the spectrum of 
$\delta\,$Ori~A.. Our interpretation of the observations is presented 
in Section 5, followed by a brief conclusion 
section.

\section{THE $\delta\,$Ori~A SYSTEM}
\label{s.delOriMain}

$\delta\,$Ori is a wide visual triple system containing three 
components: A (HD 36486), B (Bd--00$^{\circ}$.983B), and 
C (HD 36485), with apparent visual magnitudes of 
2.23$^m$, 14.0$^m$, and 6.85$^m$, respectively. Components 
Band C are at distances of 33''. and 53'', respectively, 
from the primary component, A. Our observations refer to the brightest 
component $\delta\,$Ori~A. 

\begin{table*}[!ht]
\caption{Parameters of the $\delta\,$Ori~A system}
\label{ParDelOri}
{\centering \small
\begin{tabular}{lcccl}
\hline
& \multicolumn{3}{c}{Components} & \\
\cline{2-4}
Parameter & Aa$^1$ & Aa$^2$ & Ab& References \\
\hline
Distance to the star, pc  &360  & 360        & 360   & \cite{ApelanizWalborn04} \\
Distance to component  Aa$^1$, $R_{\odot}$  &...& 33  
                               & $\approx{25\,000}$& \cite{Miller02}\\
Spectral type  &09.5II  & B0.5III    & B (early subtype)  
                                                      & \cite{Harvin02} \\

Orbital period   & ---    & $5.7325\,$ days & $\approx200\,$ yrs 
                                                      & \cite{Harvin02} \\
Radius, ($R_{\odot}$)& 11     &  4         & ...       & \cite{Harvin02,Miller02}   \\
Mass,  ($M_{\odot}$) &11.2    &  5.6       & $\approx 27$ & \cite{Harvin02} \\
Luminosity, $\log(L/L_{\odot})$& 5.26     & 4.08   &    -    & \\
$T_{eff}$, K        &33$\,$000&27$\,$000  & ...    &\cite{Voels89,Tarasov95}\\
$\lg g$             & 3.4    & 3.8        & ...    &\cite{Voels89,Tarasov95}\\
Contribution to optical radiation &70          & 7      & 23     & \cite{Harvin02}            \\
in HeI$\,\lambda\,6678\,$ line region, (\%)      &     &       &       &     \\
$V\sin i$ (km/s)    & $157\pm6$  & $138\pm16$ &$\approx 300$ &\cite{Harvin02}\\
                    & $133$  &            &              &\cite{Abt02} \\
$V_{\infty}$, km/s  & 2000   & 1500       & ...    & \cite{Lamers93,Wilson85}\\
Mass loss 
${\dot{M}},~M_{\odot}/\mbox{yr}$&$1.1\times10^{-6}$&$1.2\times10^{-7}
                                      $& ...& \cite{Lamers93,Wilson85}\\
\hline
\end{tabular}
}
\end{table*}

$\delta\,$Ori~A (HD 36486, HR 1852) is a physical triple 
system with the primary $\delta\,$Ori~Aa, which is an eclipsing 
binary with orbital period P =5.73$^d$~ \cite{Hoffleit96}, and the 
secondary $\delta\,$Ori~Ab, with an orbital period of 224.5~yrs. 
Recent detailed Doppler tomography studies~\cite{Harvin02} 
have made it possible to refine the parameters of 
the .Ori A triple system, presented in the Table.

The brightest component, $\delta\,$Ori~Aa$^1$, is a high-mass O9.5II star 
with an intense stellar wind; the star's 
mass-loss rate is $\log(\dot{M})\approx -6.0$ and the 
limiting velocity of the wind is $V_{\infty}\approx 2000\,$km/s.
A value a factor of two lower is given in~\cite{Bieging89}: 
$\lg(\dot{M})\approx -6.3$. According to~\cite{Grady84,Voels89}, 
$V_{\infty}\approx2300\,$ª¬/á. The rotational velocities of the star 
determined by different authors differ appreciably. 
In~ \cite{Harvin02}, a relatively high rotational velocity for the 
primary was obtained, $V\sin i =157\pm 6\,$km/s,, 
while Abt et al.~\cite{Abt02} found $V\sin i=133\,$km/s (see the Table). 

The study~\cite{Voels89} presents a photospheric radius for 
the primary, Aa$^1$ $R_* =22R_{\cdot}$,, twice the value 
$R_* =11R_{\cdot}$, given in the Table. This discrepancy is probably 
due to the fact that a larger distance to the star was 
used in~\cite{Voels89} than the distance in the Table, 450 pc. 

The component~$\delta\,$Ori~Aa$^2$ is a B0.5\,III star. Preliminary 
studies indicate that the third component of the system, $\delta\,$Ori~Ab, 
is an early B-subtype star with broad spectral lines, indicating either 
a high rotational velocity ($V\sin i\approx 300\,$km/s) or binarity of 
the star~\cite{Harvin02}.

Harvin et al.~ \cite{Harvin02} determined the orbital inclination 
of the $\delta\,$Ori~Aa close binary to be $i\approx 67^o$ and estimated 
the mass of the components to be M(Aa$^1)=11.2M_{\odot}$ 
and M(Aa$^2)=5.6M_{\odot}$. The secondary, $\delta\,$Ori~Aa$^2$, is 
appreciably weaker than the primary (see Table). Both the primary and 
secondary, Aa$^1$  and Aa$^2$, have masses roughly half the mass of 
main-sequence stars with corresponding positions on the HR 
diagram~\cite{Harvin02}.. It was suggested in~\cite{Harvin02}. that, in the 
course of its evolution, the system underwent a common-
envelope stage in which both stars filled their Roche 
lobe and with intense mass loss. After the system 
had lost half its total mass, the distance between the 
components increased and the mass loss ceased.

The mass of the third component of the system, Ab, cannot be derived from 
the radial velocities. The Table presents the value
$M(Ab) \approx 27\,M_{\odot}$ from~\cite{Harvin02}, which corresponds to a 
main-sequence star with $M_{\mbox{\scriptsize bol}}=-4^{m}.2$. A main-sequence 
star with a mass of $27\,M_{\odot}$ should have spectral type 
O8.5V\cite{HowarthPrinja89}, which contradicts the spectral classification 
in the Table. To explain this inconsistency, Harvin et al.~\cite{Harvin02} 
suggest that Ab may be a close binary consisting of two main-
sequence B0.5V stars with a mass of $\approx 19\,M_{\odot}$. 

\section{OBSERVATIONS AND REDUCTION OF THE SPECTRA}
\label{s.Obs}

Our spectral observations of the $\delta\,$Ori~A system were carried out 
on January 10/11, 2004 with the 6m telescope of the Special Astrophysical 
Observatory as part of our program to search for rapid line-profile 
variability in the spectra of early-type stars~\cite{PaperI}. We 
used the NES quartz echelle spectrograph~\cite{Panchuk02}, which 
is permanently mounted at the Nasmyth focus and equipped with the 
Uppsala 2048 x 2048 CCD. To increase the spectrograph's limiting magnitude, 
we used a three-cut image cutter~\cite{Panchuk03}. The resulting 
spectral resolution was $R~\approx~60000$ and the dispersion 0.033 A/pixel. 
The image size during the observations was about 3''. The reference spectrum 
was provided by a ThAr lamp. 

The angular separation between the most distant 
components, $\delta\,$Ori~Aa$^1$ and Ab, is about 0.3''. (see 
Table), which means that all components contribute 
to the resulting spectrum of the $\delta\,$Ori~A system. Over 
a total observation time of $\approx2^h50^m$, we obtained 40 
spectra of the star with an exposure time of 180 s. 
Taking into account the CCD readout, the time resolution 
was 260 s. The signal-to-noise ratio per pixel .
was 500 in the blue (4500~\AA) and 800 in the red .
(6000~\AA).

The preliminary reduction of the CCD echelle spectra was done in the MIDAS 
package~\cite{PaperI}. We adapted the standard ECHELLE procedures of the 
MIDAS package for data obtained with an image cutter. The reduction stages 
included \\
1) median filtering and averaging of the bias 
frames with their subsequent subtraction from the 
remaining frames; 
2) cleaning of cosmic rays from the frames; 
3) preparation of the flat-field frame; 
4) determination of the position of the spectral 
orders (using the method of Ballester~\cite{Ballester94}); 
5) subtraction of scattered light (to determine the 
contribution of scattered light, we distinguished the 
inter-order space in the frames, and then carried out 
a two-dimensional interpolation; this function was 
recorded in individual frames to be subtracted from 
the initial images); 
6) extraction of the spectral orders from the reduced 
images of the stellar spectrum, the flat fields, 
and the wavelength reference spectra; flat-field reduction; 
7) wavelength calibration using a two-
dimensional polynomial approximation for the identifications 
of the lines of the reference spectrum in 
various spectral orders. 

When the image cutter is used, each order of 
the echelle image is represented by three sub-orders 
(cuts). In a given order, the instrumental shift of the 
upper and lower cuts relative to the middle cut was 
determined via a cross-correlation with the reference 
spectra. The three cuts obtained in a given order 
were summed using median averaging, taking into 
account the determined instrumental shifts. 

To study the line profile variability, the processed 
spectra were normalized to the continuum level reconstructed 
in each echelle order. We drew the continuum 
in spectral orders containing narrow spectral 
lines using the technique developed by Shergin et 
al.~\cite{Shergin96}, in which spectra are smoothed with a variable 
Gaussian filter with a 25-30~\AA window, with the same 
parameters for the entire set of spectra. 

In orders containing broad (usually hydrogen) 
spectral lines, the continuum was drawn using the 
following procedure. In all 40 spectra, broad lines 
were cut out in fixed wavelength intervals. To determine 
the position of the continuum, we used a 
polynomial approximation for all wavelengths of the 
order, except for the regions of the cut-out broad 
spectral lines. The approximation parameters remained 
unchanged for all spectra of the set. This 
procedure provides stability and reproducibility of our 
determination of the continuum in all the spectra 
with an accuracy of up to several tenths of a percent. 
This makes it possible to reach high accuracy in 
deriving the differential line profiles and detecting 
profile variability in broad lines at levels down to 0.1\%. 

\section{LINE PROFILE VARIABILITY} 
\label{s.ProfAnal}

\subsection{Contribution from Different Components of the System
            to the Line Profiles} 
          \label{ss.ProfMod}

     We selected sufficiently strong unblended lines for detailed studies of 
the line-profile variations. The selection criterion for absorption lines was 
that the residual intensity be $r_{\mbox{\scriptsize max}}>0.1$. These criteria 
led to the selection of the HeII$\,\lambda\,$4686, HeI$\,\lambda\,$4713 
 and  H$_{\beta}$ lines. In addition, we studied the profiles of the 
CIII$\,\lambda\,$5696 emission line. 

The $\delta\,$Ori~A system contains three O and B stars, 
each of which could have variable line profiles; the 
observed profiles result from the contributions of all 
stars. It follows from the Table that above two-thirds 
of the flux from the system at the wavelength of the 
HeI$\,\lambda\,6678\,$ line comes from the primary component, 
Aa$^1$. The contributions from Aa$^2$ and Ab at these 
wavelengths are substantially smaller. 

At other wavelengths, the contributions from Aa$^2$
and Ab may differ from those in the Table. To determine 
to what extent each of the stars in the system is 
responsible for line-profile variability, the contribution 
of each star to the total profiles of the analyzed lines 
must be clarified. In addition, taking into account the 
substantial scatter of the $V\sin i$ values for Aa$^1$ (see the 
Table), reliable v sin i estimates for the components of 
the $\delta\,$Ori~A system are highly desirable.

\begin{figure}[!ht]
\centering
\includegraphics[height=9.0cm,width=6.0cm,angle=00]{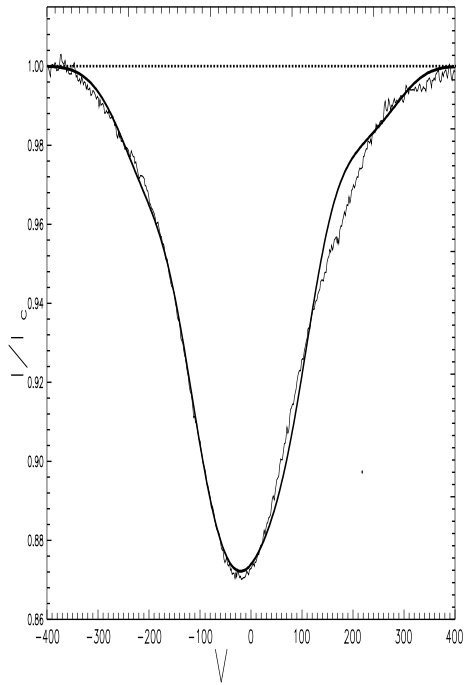}
  \caption{\footnotesize HeII.4686 synthetic (bold curve) and observed 
            (thin curve) line profiles, plotted in terms of the Doppler 
            shift V from the line center.   
          }
\label{fig.ProfModel}
\end{figure}

To solve these problems, we modeled the combined 
profiles for the photospheric HeII$\,\lambda\,$4686 and HeI$\,\lambda\,$4713  
lines in the spectrum of $\delta\,$Ori~A. We assumed 
the total line profile to be the sum of the 
contributions from the components, since Aa$^1$ and 
Aa$^2$ are not eclipsed at the above orbital phases. We 
calculated the rotation-broadened line profiles using 
the standard relations (see, for example,\cite{Gray80}). The 
radial velocities of the components for the time of the 
observations were calculated using the ephemeridas of 
Harvin et al.~\cite{Harvin02}. 

During our observations, the orbital phase varied 
in the narrow interval $0.848 - 0.851$. For such small 
phase variations, the velocities of the components 
can be considered to be constant over the total observation 
time. The heliocentric velocities $V(\mbox{Aa$^1$})=-21\,$km/s, 
and $V(\mbox{Aa$^2$})=-102\,$km/s were obtained 
using the ephemeris~\cite{Harvin02}. 

For convenience in comparison, both the calculated 
and observed line profiles were transformed into 
the system of the center of mass of the close binary 
$\delta\,$Ori~Aa. The velocity
$V_{\mbox{\scriptsize r}}=\approx 23.0\,$ª¬/á~\cite{Harvin02} was used 
as the zero-point of the velocity scale. In this system, 
the velocity of the third component is 
$V(\mbox{Ab})=\approx27\,$km/s~\cite{Harvin02}. 

Figure 1 presents the total profile of the photospheric 
HeII$\,\lambda\,4686$ line calculated using these parameters. 
The observed profiles are reproduced very 
reasonably. The small discrepancies between the observed 
and calculated profiles near $V=150-250\,$km/s 
may be due to the contribution of the stellar wind from Aa$^1$.

\subsection{Variations of the Average Profiles. Differential Profiles}
     \label{s.LPV} 

The nightly-average line profiles in O-star spectra 
are often substantially variable~\cite{deJong01}. Line-profile 
variations on shorter time intervals are insignificant. 
To make the line-profile variations in the spectra of 
$\delta\,$Ori~A in time intervals of 30-40 min more clearly 
visible, we divided the profiles into four groups with 
ten profiles in each and calculated the average profiles 
for these groups. 

\begin{figure*}[!ht]
\centering
  \includegraphics[height=18.0cm,width=15.0cm,angle=00]{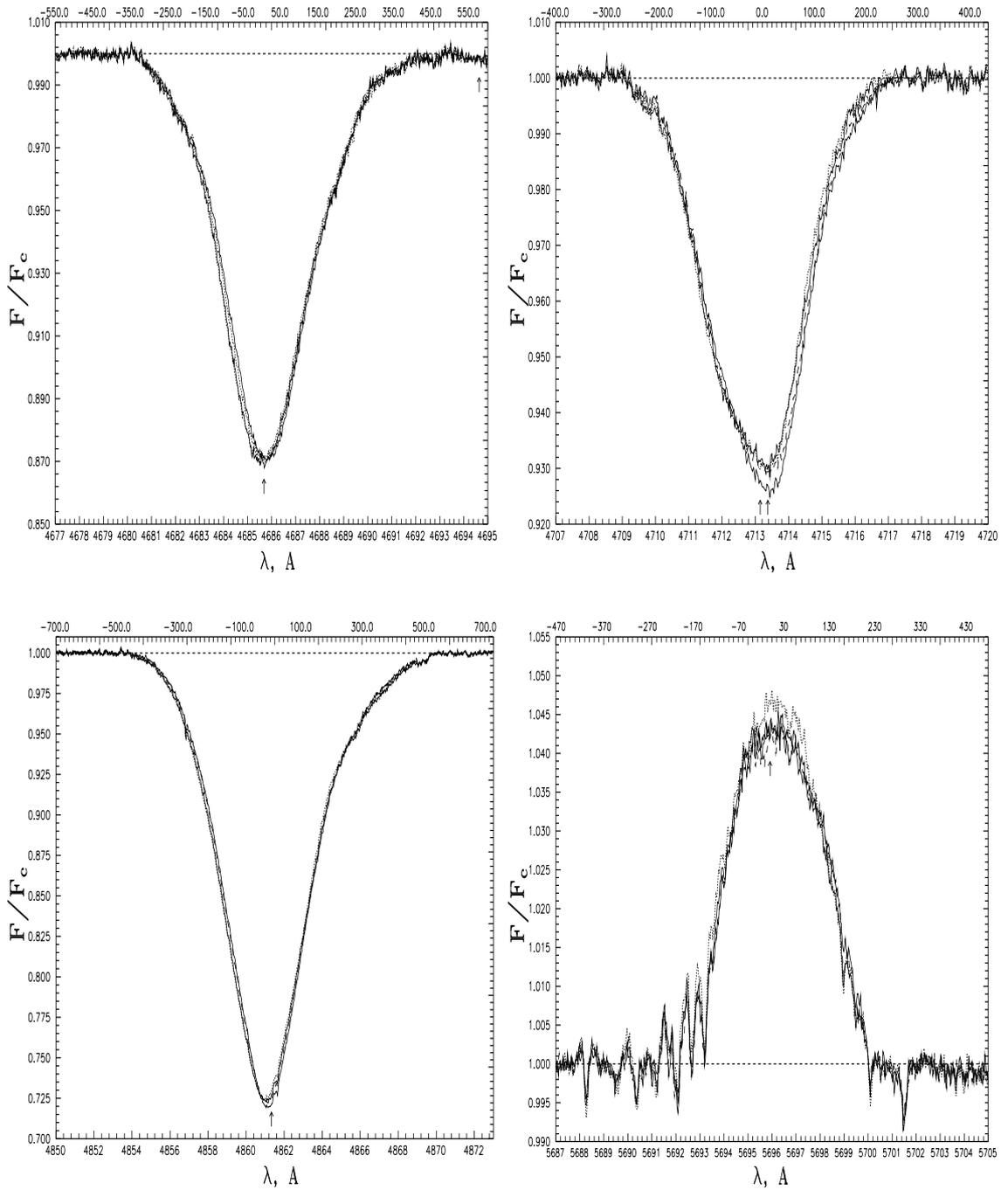}
\caption{\footnotesize Average profiles for spectra 1-10 (thin solid curve), 
         11-20 (dashed curve), 21-30 (bold solid curve), and 31-40 (dotted 
         curve) for the HeII$\,\lambda\,$4686 (top left), HeI$\,\lambda\,$4713 
         (top right),  H$_{\beta}$ (bottom left), and CIII$\,\lambda\,$5696 
         (bottom right) lines. The flux in 
         the line frequencies, $F$, is given as a fraction of the flux in 
         the continuum adjacent to the lines, $F_{\rm c}$. The arrows 
         indicate the laboratory wavelengths of the lines. 
         }
\label{fig.MeanGroup}
\end{figure*}

The average profiles for each group presented in 
Fig.~\ref{fig.MeanGroup} display appreciable variations within an hour. 
These variations are particularly clearly visible in the 
HeI$\,\lambda\,$4713. In the transition from the first group 
of spectra to the following three groups, the line depth 
decreases by 1\%-1.5\%. These variations are correlated 
with variations in the groups of average profiles 
for the CIII$\,\lambda\,$5696 emission line. In the transition from 
the first to the fourth group of profiles, the flux in the 
central region of the line increases by $\approx 1\%$.The amplitude 
of the flux variations in the H$_{\beta}$ and HeII$\,\lambda\,$4686  
lines reaches 1\%-2\%.

In the blue wing of the CIII$\,\lambda\,$5696 profile, narrow absorption 
features with depths less than 0.01 of the continuum intensity can be seen. 
Similar, but less deep, features are also present in the red wing of the 
line. Our analysis shows that the positions of all of 
these features coincide with those for atmospheric 
absorption lines, primarily water vapor~\cite{Pierce73}.

To reveal variable features in the line profiles, we 
constructed differential profiles by subtracting the average 
profile obtained using all forty available spectra 
of $\delta\,$Ori~A from the individual profiles.

\begin{figure*}[!ht]
 \centering
  \includegraphics[height=15.0cm,width=12.0cm,angle=00]{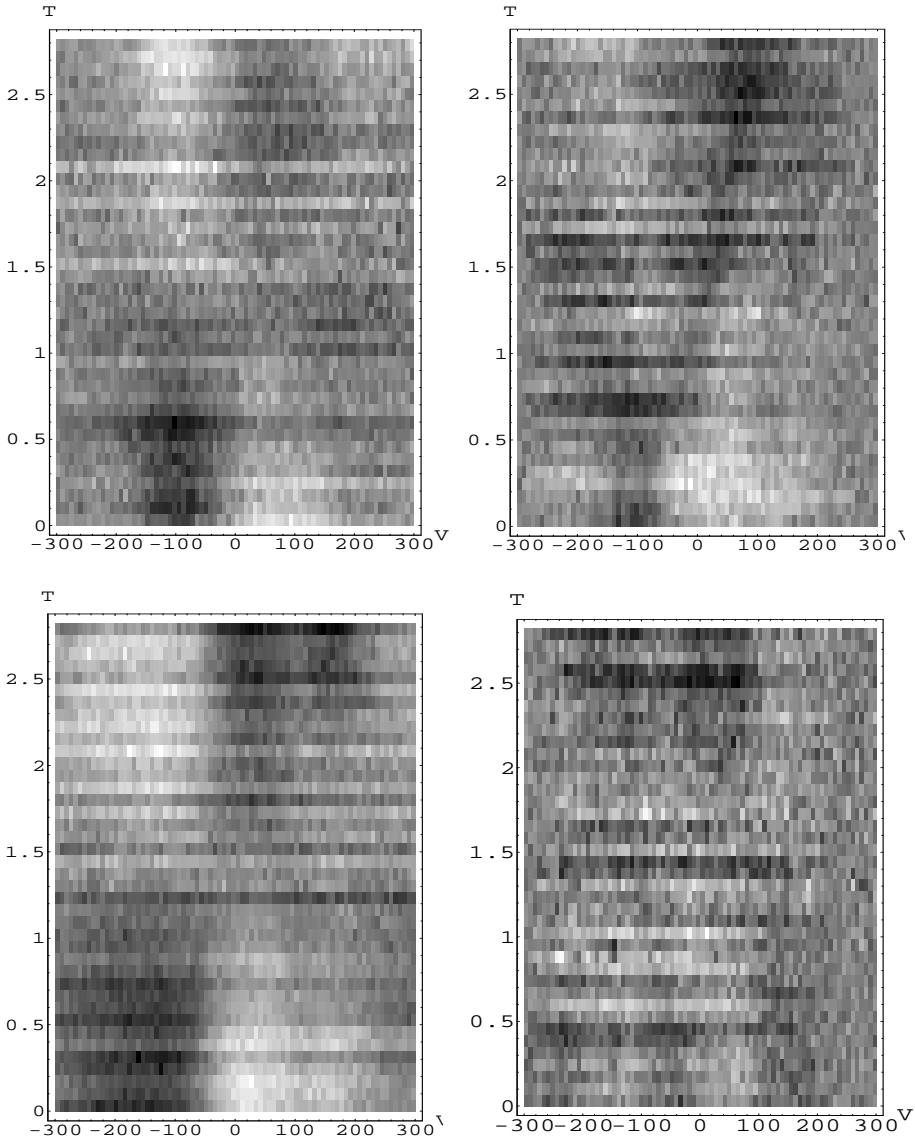}
\caption{\footnotesize Dynamical spectra for the profile variations for the 
           HeII$\,\lambda\,$4686 (top left), HeI$\,\lambda\,$4713 (top right), 
           H$_{\beta}$ (bottom left), and CIII$\,\lambda\,$5696 (bottom right) 
           lines. The interval between consecutive spectra is four minutes. 
           Dark regions in the diagrams correspond to intervals brighter then 
           the average profile (peaks), and light regions to less bright 
           intervals (valleys). 
        }
\label{fig.LPV}
\end{figure*}

Figure~\ref{fig.LPV} presents dynamical differential profiles 
of the studied lines in the spectrum of $\delta\,$Ori. Grey 
shades indicate deviations of individual profiles from 
those averaged over the entire set of spectra. For 
convenience, the wavelength scale is translated into 
Doppler shifts from the line center. The radial velocity 
of the center of mass of the $\delta\,$Ori~A system was used 
as the zero point of the wavelength scale (see Section~\ref{ss.ProfMod}).

We show the differential profiles in the "negative", i.e., darker regions in 
Fig.\ref{fig.LPV}correspond to regions of the profile that lie above the 
average value (peaks), while lighter regions correspond to parts of the 
profile that lie below the average value (valleys). This makes 
it easier to identify regular variations of the profiles 
in Fig.\ref{fig.LPV}. The HeII$\,\lambda\,$4686, HeI$\,\lambda\,$4713, 
and H$_{\beta}$ line profiles show that the peaks move from the violet to 
the red wing of the profile.

We can see a broad (50-100 km/s), variable feature in the vicinity of the 
HeII$\,\lambda\,$4686 line. The feature appears at a velocity near 
-100 km/s in the first (lower) differential profiles, and is then gradually 
shifted towards the line center. In the last (upper) 
profiles, it is visible in the domain of positive velocities. 
Such behavior of features in the differential profiles 
is typical for non-radial pulsations~\cite{TS97II}. Virtually no 
obvious spectrum-to-spectrum profile variations near weaker spectral lines, 
due to both the short duration of the observations and the low amplitude of 
the profile variations. We used the technique described in the following 
subsection to reveal profile variations in such 
weak lines.

\subsection{Analysis of the Spectrum of Time Variations of the
            Differential Spectra
            }

The method of Fullerton et al.~\cite{Fullerton96} is often used 
to elucidate the presence of weak line profile variability. 
We describe here a substantially modified version 
of this technique, which we used in our analysis. 
Suppose that N spectra of a studied star have been 
obtained. We denote $F_i(\lambda),~i=1, ... N$ to be the 
flux in the ith spectrum at wavelength ., normalized 
to the continuum. Let $\overline{F}_i(\lambda)$ be the flux at wavelength 
$\lambda$ averaged over all the observations. 
The dispersion of the random ${F}_i(\lambda)$ is then equal to 
%
\begin{eqnarray}
\label{Eq.VarFlmbd}
&D(\lambda)= TVS(\lambda)&\\ \nonumber
&=\frac{1}{N-1} 
\left(
  \sum\limits_{i=1}^{n}g_i\left[F_i(\lambda)-
                  \overline{F}_i(\lambda)\right]^2
     \left/{\sum\limits_{i=1}^{n}g_i}\right.
\right) \, & .
\end{eqnarray}

Here, $g_i$ is the relative weight of the $i$th observation, 
which is inversely proportional to $\sigma_{i}^{c}$ -- the standard  
deviation $\overline{F}_i(\lambda)$ at wavelengths near the line whose 
profile variability is being studied. This definition for 
$g_i$ makes it possible to take into account the different 
qualities of the data in the analyzed set of spectra. 
$TVS(\lambda)$, or the Time Variation Spectrum~\cite{Fullerton96}, obeys 
a $\chi^{2}/(N-1)$ distribution with $N-1$ degrees of freedom. 

In the case of profile variability with a sufficiently high amplitude, the 
$TVS(\lambda)$ in the vicinity of the corresponding line substantially 
exceeds its values in the adjacent continuum. To ensure that the increase 
in the amplitude is due to real variability, we specified a small significance 
level $\alpha \ll 1$ for the hypothesis that the excess is due to a random 
variation of the noise component of the profile. The quantites for the 
$\chi^{2}$ distribution are calculated in the ordinary way for this 
$\alpha \ll 1$ level (see, for example,~\cite{Brandt}). 
Let $\chi^{2}_{\alpha}$ be specified so that the probability 
$P(\chi^{2}/(N-1) > \chi^{2}_{\alpha}) =\alpha$ 
If the calculated $TVS(\lambda)$ value exceeds $\chi^{2}_{\alpha}$, 
the hypothesis that the line profile is variable can be adopted.

As was noted in~\cite{PaperI}, for lines with small-amplitude profile 
variations or with a small number of spectra, the function $TVS(\lambda)$ 
cannot be used to prove whether a given line profile is variable. In the 
absence of any visible variations, we determined whether a 
line profile was indeed variable using the following procedure.

Before the standard-deviation spectrum was obtained, the differential spectra 
were pre-smoothed using a filter that is broad compared to the pixel size. 
If the filter width is $\Delta\lambda$ on the wavelength scale, 
then the amplitude of the smoothed random (noise) component of the 
differential profiles $\overline{N_j(t)}$  decreases after smoothing by a 
factor of $\approx \sqrt{\Delta\lambda/\delta\lambda}$, where $\delta\lambda$ 
is the width of an individual pixel on the wavelength 
scale in the vicinity of the line. If the width of the 
variable component is not smaller than that of the 
filter, then smoothing with this filter will not substantially 
change the amplitude of the variable component, 
while, after the smoothing, peaks in the standard 
deviation spectrum that correspond to the variable 
component can be detected with increased reliability.

It was shown empirically that the best results are obtained for smoothing with a 
Gaussian filter with a width of $S=0.5 - 1.0$ \AA~(15-30 pix.). To more clearly 
demonstrate the method used to search for weak profile variations, we used a 
smoothed time-variation spectrum, smTVS, which is a collection of spectra of 
the normalized standard deviation $\sigma(\lambda,S)=\sqrt{TVS(\lambda,S)}$  
with the differential spectra smoothed with a Gaussian filter with a variable 
width $S$. 

\begin{figure}[!ht]
 \centering
  \includegraphics[height=8.5cm,width=8.2cm,angle=00]{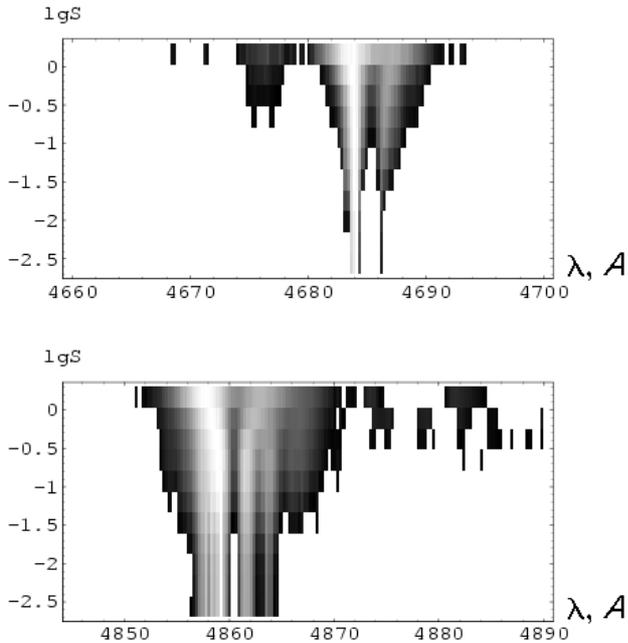}
\caption{\footnotesize Time variation spectrum for the HeII$\,\lambda\,$4686 
          (top) and H$_{\beta}$ (bottom) line profiles. The vertical axis 
          plots the logarithm of the filter width in \AA. 
        }
\label{fig.smTVSdelOri}
\end{figure}

Figure~\ref{fig.smTVSdelOri} presents density diagrams for the time-variation 
spectra for the HeII$\,\lambda\,$4686 and H$_{\beta}$ line profiles. Darker 
areas correspond to higher amplitudes of the smTVS spectrum. Only values 
corresponding to significance levels $\alpha < 10^{-4}$ for the hypothesis that
the profiles are not variable are presented. The density diagrams show that 
the variability of the H$_{\beta}$ and HeII$\,\lambda\,$4686 lines is clearly 
present.

The smTVS spectrum for the HeII$\,\lambda\,$4686 line profile with the filter 
width $S>0.3\,$\AA indicates variability of the 
CIII$\,\lambda\,$4673.95, OII$\,\lambda\,$4673.75, 4676.234, 
NII$\,\lambda\,$4674.909, 4678.14, and SiIII$\,\lambda\,$4673.273 lines, 
which cannot be detected using ordinary methods. 
With larger filter widths, variability of a large group 
of OII, NIII, and ArII lines at wavelengths from.
4871 to 4890 \AA becomes visible. Note that, in spite 
of the fact that these lines are extremely weak and 
their residual intensities do not exceed the noise level (by one pixel along the 
spectrum) at the adjacent continuum, their variability is clearly detected. 
Unfortunately, our technique cannot be used to localize weak lines with 
variable profiles accurately when larger filter widths are used. 

Note that the efficiency of this method of using the 
smTVS spectra to search for weak line-profile variability 
is sensitive to the number of spectra obtained, 
and increases substantially as this number increases. 

\subsection{Wavelet Analysis for the Line-Profile Variations} 

Analysis of the differential line profiles in the spectra of a star 
(Fig.~\ref{fig.LPV}) indicates the presence of a number of discrete features. 
Many small-scale features are connected with the noise component of the 
profiles, whereas larger-scale features may be due to the regular 
component of the profile variations\cite{PaperI}. A very suitable mathematical 
technique for studying the development of different-scale profile features is 
wavelet analysis (see, for example,\cite{Ast96,LM99}). It is advisable 
to use as the analyzing (or mother) wavelet the so-
called MHAT wavelet,
$
\psi(x)\!\!=\!\!(\!1-x^2)\!\exp(\!-x^2/2)\, ,
$
which displays a narrow energy spectrum and whose zero and first momenta are 
equal to zero. The MHAT wavelet is essentially the second derivative of a 
Gaussian function (taken with a minus sign) that can be 
used to describe features in the differential line profiles.

Using this wavelet, the integral wavelet transformation 
can be written\cite{Ast96,KH03}: 
%
\begin{eqnarray} \label{Wab}
  &  W(s,u)=\frac{1}{s}\int\limits_{-\infty}^{\infty}f(x)
    \psi\left(\frac{x-u}{s}\right)dx & \\ \nonumber 
  & = \, \int\limits_{-\infty}^{\infty}f(x) \,
 \psi_{su}(x)dx & \,. \qquad
 \end{eqnarray}
%
where $f(x)$ is the studied function (in our case, a 
differential line profile). The signal energy energy, 
$E_W(s,u)\!\! = \!\!W^2(s,u)\!$, characterizes the energy distribution 
of the studied signal in $(s,u)=\,$= (scale, coordinate) 
space. Further, we will analyze the differential 
line profiles in the form $r(V)=F_i(V)-\overline{F}_i(V)$,where the Doppler 
shift $V$ from the central frequency of the line is used as the x coordinate. 
In this case, the scale variable $s$ is expressed in km/s.

To study the evolution of features in the differential 
profiles, we calculated wavelet spectra, $W(s,V)=W(s,V,t)$, for the 
HeII$\,\lambda\,$4686, HeI$\,\lambda\,$4713, H$_{\beta}$ and the 
CIII$\,\lambda\,$5696 emission line for all times $t$ corresponding to our 
spectra. We will call the collection of functions $W(s,V,t)$ the dynamical 
wavelet spectrum for the differential line profiles. 

\begin{figure*}[!ht]
\centering
\includegraphics[height=20.0cm,width=16.5cm,angle=00]{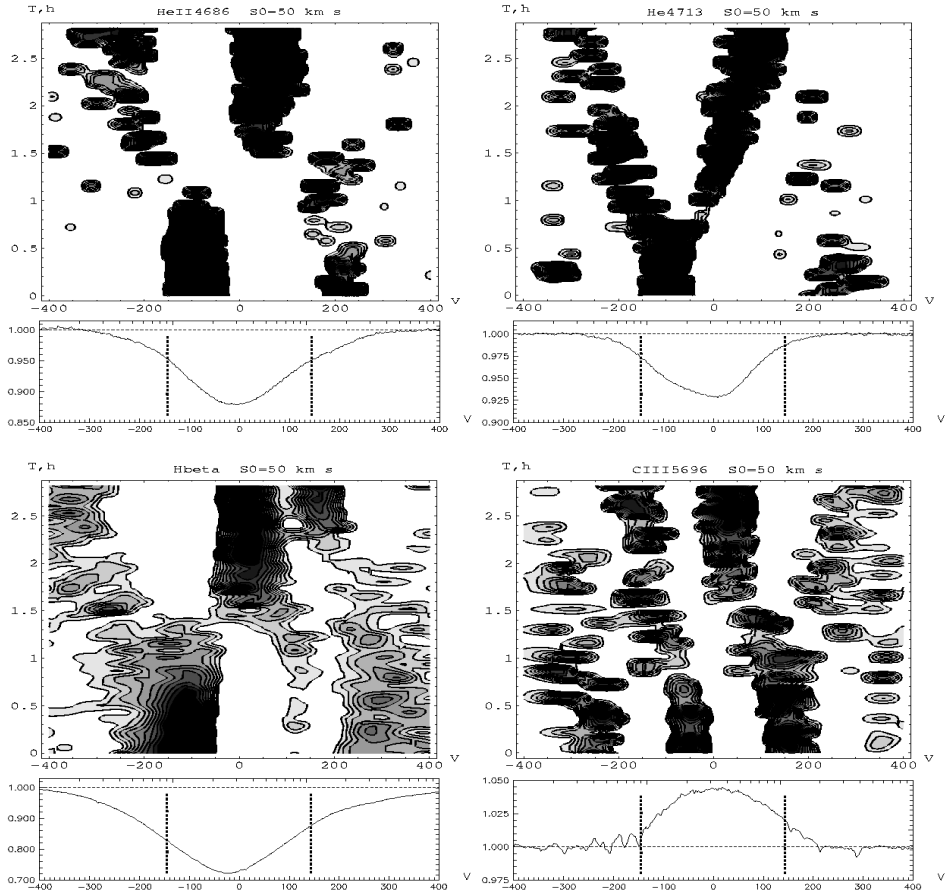}
\caption{\footnotesize Dynamical wavelet spectrum of the 
           HeII$\,\lambda\,$4686 (top left), HeI$\,\lambda\,$4713 (top right), 
           H$_{beta}$ (bottom left), and CIII$\,\lambda\,$5696 (bottom right) 
           line-profile variations for the scale parameter $S_{0}$=50 km/s. 
           The corresponding line profiles averaged over the entire set of 
           observations are given beneath the wavelet spectra. The dotted 
           curve indicates the region $\pm V\sin i$ for the primary 
           component Aa$^1$ of the $\delta\,$Ori~A system. 
         }
\label{fig.dynWL50}
\end{figure*}

Figure~\ref{fig.dynWL50} presents density diagrams for the dynamical wavelet 
spectra of the studied lines for the scale parameter $s=50\,$km/s. The diagrams
present the ratios $d=(W-W_{min})/\Delta W$ rather than the function 
$W(s,V,t)$ itself. Here, $\Delta W=W_{max}-W_{min}$, where  where $W_{max}$ 
and $W_{min}$ are the maximum and minimum of $W(s,V,t)$ for all possible 
$t$ in the interval of $V$ within the line profile. Darker regions 
correspond to larger values of $W(s,V,t)$. We have marked only the values for 
the parameter $d\ge d_{cut}$ -- the cutting parameter of the 
wavelet spectrum. In Fig.~\ref{fig.dynWL50} dcut =0.6. 

For a comparison, the average line profiles are placed beneath the wavelet 
spectra diagrams. In the average profiles, the dotted lines delineate the 
region $-V\sin i \div V\sin i$ for the primary, Aa$^1$  (see Table).

We can see in Fig.~\ref{fig.dynWL50} that all the wavelet spectra display 
broad structures moving along the line profiles, with features moving from 
the violet to red edge of the line being most prominent. The same 
type of features are visible in the wavelet spectra of the absorption 
HeII$\,\lambda\,$4686, HeI$\,\lambda\,$4713 and H$_{\beta}$ differential line 
profiles, moving from $-110 \div -100\,$km/s to $\approx 100 \div 110\,$km/s 
over the total observation time, $2^h\,50^m$. The estimated crossing time 
$T_{\mbox{\scriptsize{cross}}}$ for the entire band of $-V\sin i \div V\sin i$ 
is $\approx 4^h$. Note the similarity between the dynamical wavelet spectrum 
and the time variation spectrum for the line profiles (see 
Section~\ref{s.LPV}).

The HeI$\,\lambda\,4713$ and HeII$\,\lambda\,4686$ lines are basically formed 
in the photosphere, and their dynamical wavelet spectra behave in a very 
similar way. Apart from the primary, which shifts towards the 
red part of the line profile, another, weaker, component 
is seen. This secondary component appears $\approx 40$~min after the start of 
the observations at $\approx -130\div\-150\,$km/s and moves towards more 
negative velocities ($\approx -300\,$300km/s) by the end of the observations. 
This feature may correspond to component Ab of the $\delta\,$Ori~A system, 
since profile variations associated with possible pulsations of this star 
should be traced up to $V=V_{min}\approx -300\,$km/s (see the Table). It is
also possible that this feature is formed in the powerful stellar wind from 
the primary of the $\delta\,$Ori~A system. 

The same secondary component as that seen for the HeI$\,\lambda\,4713$ and 
HeII$\,\lambda\,4686$ lines (although less distinctly pronounced) is also 
present in the dynamical wavelet spectrum of the H$_{\beta}$ line. Appreciable 
features of the dynamical wavelet spectrum of H$_{\beta}$ are visible even 
beyond the $\pm V\sin i\approx 300\,$km/s band for the most rapidly rotating star, 
Ab, probably indicating that the intense stellar wind from the primary 
contributes substantially to the H$_{\beta}$ line profile.

The dynamical wavelet spectrum of the differential we can see two main 
features, moving almost parallel profiles of the CIII$\,\lambda\,5696$ 
emission line differs appreciably from those for the absorption lines.
In Fig.~\ref{fig.dynWL50} we cam see two main features, moving almost parallel
in the $-200 \div 200\,$km/s band from the red to the violet 
part of the line profile. 

\begin{figure*}[!ht]
\centering 
\includegraphics[height=20.0cm,width=16.5cm,angle=00]{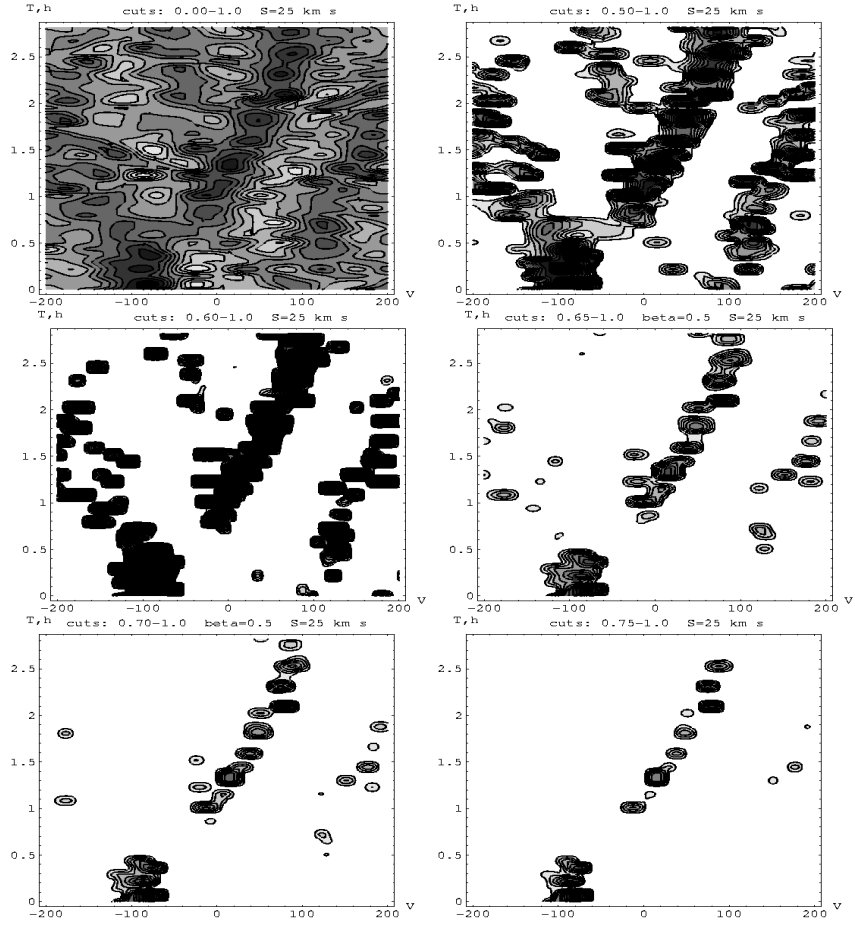}
\caption{\footnotesize Dynamical wavelet spectrum of the HeI$\,\lambda\,$4713 
                       line profile variations for $s_0$ =25 km/s and 
                       various cutting parameters $d_cut$ = (a) 0, (b) 0.5, 
                       (c) 0.6, (d) 0.65, (e) 0.7, (f) 0.75.   
        }
\label{fig.dynWL4713}
\end{figure*}

Regular structures in the wavelet spectra are most clearly visible in the  
profile of the HeI.4713 line. To study these structures in detail, 
Fig.~\ref{fig.dynWL4713} presents the calculated dynamical wavelet spectrum 
for this line for the wavelet-spectrum cutting parameters $d_{cut}$ = 
0, 0.5, 0.6, 0.65, 0.7, 0,75 and for $S_0=25\,$km/s. 
As $d_{cut}$ increases, more pronounced structures of the 
wavelet spectrum become more dominant. All the 
wavelet spectra display movement of a feature from 
-120 to 90 km/s; for $d_{cut} \ge 0.75$, only this main 
feature is present in the wavelet spectrum. 

To clarify the nature of the features revealed in the 
wavelet spectra for the differential profiles, we must 
determine whether the line profile variations related to 
these features are regular. We consider this question 
in the following section. 

\subsection{Search for Regular Variability}

To search for periodic profile variations, we carried out a Fourier analysis 
of the line-profile variability in the spectrum of $\delta\,$Ori~A. The 
Fourier spectra were cleaned of false peaks using the CLEAN~\cite{Roberts87} 
procedure modified by Vityazev~\cite{Vyt01b}. 

For each of the studied lines, we constructed the time series 
$\Delta F(t,\lambda)$ of differential fluxes for a given profile at time $t$ 
for a given $\lambda$. For convenience, the $\lambda$ values were 
converted into Doppler shifts $V$ from the profile center in the center-of-mass
frame of the $\delta\,$Ori~A triple system (see Section~\ref{ss.ProfMod}). 

This CLEAN procedure was applied to all the observed $V$ values along a line 
profile. To smooth the noise component of the profile variations, the 
$\Delta F$ values were averaged within a spectral window with width 
$\Delta V$ = 5-6~km/s. Our calculations showed that our specific selection 
of $\Delta V$ do not appreciably influence the resulting Fourier spectra. 

\begin{figure*}[t!]
\centering
\includegraphics[height=16.0cm,width=13.5cm,angle=00]{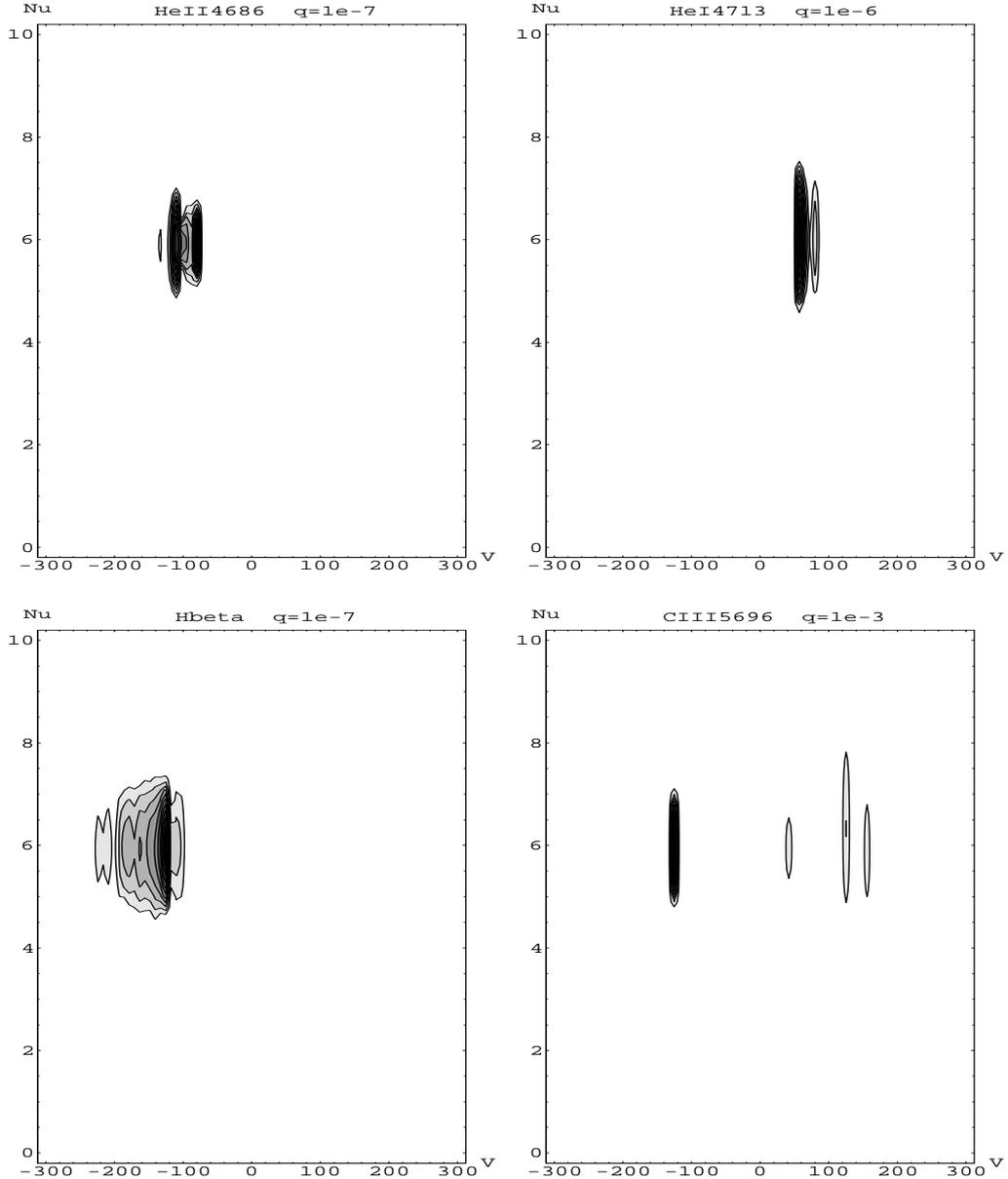}
\caption{\footnotesize Fourier power spectra for the line profile variations. 
           Fourier spectral amplitudes are presented for low significance 
           levels for the hypothesis that there is a strong peak in the white 
           noise periodogram $q$ ,for the HeII$\,\lambda\,$4686 (top left; 
            $q= 10^{-7}$), HeI$\,\lambda\,$4713 (top right; $q= 10^{-6}$), 
            H$_{\beta}$ (bottom left; $q= 10^{-7}$), and 
            CIII$\,\lambda\,$5696 (bottom right; $q= 10^{-6}$) lines. 
           Darker regions correspond to larger amplitudes of the Fourier 
          spectra. 
        }
\label{fig.FSpTot}
\end{figure*}

Figure~\ref{fig.FSpTot} presents contour plots of the square of the 
amplitude of the Fourier transform (Fourier spectra) for the 
HeII$\,\lambda\,$4686,  HeI$\,\lambda\,$4713, CIII$\,\lambda\,$5696 and 
CIII$\,\lambda\,$5696 lines. When constructing the density diagram for the 
power spectrum of the Fourier transformation, all values of the periodogram 
corresponding to low significance levels $q=10^{-7}-10^{-3}$  for the 
hypothesis that there was a strong white noise peak in the periodogram 
were rejected. Thus, only values corresponding to significances for the 
hypothesis that there is a harmonic component in the periodogram for 
a value $\alpha=1-q>0.999$ are presented in the plots. 

Figure 7 has a broad peak at the frequency $\nu=5.9\pm 1\,d^{-1}$. The large 
width of the peak is due to the relatively low resolution of the Fourier 
spectrum, related to the short duration of the observations. The 
value $P=1/\nu \approx 4^h.1\pm 0^h.7$ exceeds the duration of 
the observations, and so cannot be directly identified 
with the period of the regular profile variations. We 
will call this value the quasi-period for the regular 
profile variations. Of course, the presence of regular 
line-profile variations with the proposed period 
must be confirmed by more extended observations. At 
present, we can only state that variations of the differential 
profiles in the spectrum of $\delta\,$Ori~A are consistent with the 
period $\approx 4.1\pm 0.7^h$, with the same period being indicated for all 
the studied lines. 

\section{DISCUSSION}
\label{s.Disc}

Our analysis shows that regular line profile variations with the characteristic
time $P=4^h-6^h$ may occur in the spectrum of the $\delta\,$Ori~A system. 
Such periods are typical for profile variations due to non-radial pulsations 
(NRPs) in O stars~\cite{deJong99}.

Apart from NRPs another possible origin for the rapid line profile variations 
is rotational modulation of the profiles. In the model of Kaper 
et al.~\cite{Kaper97}, it is assumed that, along with a spherically symmetrical
outflow in the form of a stellar wind, some denser, compact corotational streams 
rotating with the angular velocity of the star could be present in the 
atmosphere. It is proposed that additional absorption of the stellar 
radiation when these streams cross the line of sight results 
in regular line profile variations. The period of these 
variations will be $P_{n} = P_{rot}/n$, where $n$ is the number 
of corotational streams. 

The orbital period $P_{rot}$ of the primary Aa$^1$ can be estimated from the 
data in the Table: $P_{\mbox{\scriptsize rot}}/\sin i \approx 3.6\,$d. 
The real inclination $i$ of the rotation 
axis of the primary is probably close to the orbit inclination, 
$67-77^{o}$~\cite{Harvin02}, which corresponds to $P_{rot}/sini \approx 3.6$~d.
Explaining the quasi-period of the variations $P\approx4^h.1$ as a result of 
rotational modulation of the profiles requires the value $n =21$, which does 
not seem plausible. In addition, in this case, we should see no fewer 
than 10 regular components in the entire line profile, 
which are not observed (Fig.~\ref{fig.LPV}). 

Based on this reasoning, we conclude that the rotation of Aa$^1$ cannot give 
rise to the detected rapid profile variations with the quasi-period 
$P\approx4^h.1$. Similar estimates based on data from the Table 
indicate that the rotation of Aa$^2$ and Ab likewise 
cannot explain the profile variations in the spectrum of $\delta\,$Ori~A.

Thus, the most likely origin of the regular profile variations is photospheric 
NRPs of the stars in the $\delta\,$Ori~A system. The relations 
$l \approx 0.1 + 1.09 \mid\Delta\phi_0\mid/\pi$ and  
$m \approx \-1.33 + 0.54 \mid \Delta\phi_1\mid/2pi$ 
can be used to determine the (l m) mode of the NRPs, where $\Delta\phi_0$ 
is the phase difference of the Fourier components of the line-profile 
variations of the main component $\nu_0$ of the NRPs and $\Delta\phi_1$ is the 
same value for its first harmonic $\nu_1=2\nu_0$~\cite{TS97II}. 
Unfortunately, due to the short duration of the data set 
$\Delta\phi_0$and $\Delta\phi_1$ cannot be determined with sufficient 
accuracy. The value $\Delta\phi_0$ for the H$_{\beta}$ line profile can be 
determined only in the velocity interval $-18 \div 150\,$km/s, and is 
equal to $\approx \pi$. Assuming that this value can be extrapolated to the 
entire velocity interval $\pm V\sin i$,we find $l \approx 2$. 

Note that the profile variations of the studied lines in the spectrum of 
$\delta\,$Ori~A are qualitatively similar to those in the spectra of numerous 
Be stars~\cite{RiviniusEtAl03} that can be described as the result of NRPs 
in the sector mode $(l,\mid m\mid)=(2,2)$.

Let us now analyze the line profile variations in the spectrum of 
$\delta\,$Ori~A supposing that they are related to NRPs in the 
$(2,\pm 2)$ mode. The velocity of a volume element of the star undergoing 
NRPs in spherical coordinates is $\propto Y_{ml}(\theta,\phi)exp(i\omega t)$, 
where $Y_{22}(\theta,\phi) \propto (1-\cos2\theta)*exp(im \phi)$ is a 
spherical harmonic and $\omega$ is the angular frequency of the 
pulsations~\cite{TS97II}. 
The time dependence of the velocity is specified by  the factor 
$\exp(im \phi + \omega t)$, where $2\pi/\omega$ is the pulsation period. The 
phase velocity for the propagation of perturbations across the stellar surface 
is $\omega_{NRP} = 2\pi*\omega/m$. This relation implies that the perturbation 
propagation period is twice the pulsation period.

The propagation of perturbations of the velocity of 
matter in the photosphere corresponds to movement 
of features in the differential profiles (peaks and valleys) 
with the same periods. This means that a profile 
feature associated with NRPs will cross the band 
$\pm V\sin i$ over for the time $T_{NRP}$, where $T_{NRP}$ is the NRP period. 
Therefore, we can write the approximate equality 
$T_{NRP} \approx T_{\mbox{\scriptsize{cross}}}$, where 
$T_{\mbox{\scriptsize{cross}}}$ is the time for the feature to cross the 
band $\pm V\sin i$ in the differential profile. We found the time 
$T_{\mbox{\scriptsize{cross}}}$ from our analysis of the wavelet spectra 
of differential profiles of the HeII$\lambda$4686 and HeI$\lambda$4713 
photospheric lines (Fig.~\ref{fig.dynWL50}):
$T_{\mbox{\scriptsize{cross}}} \approx 3-4^h$, very close to the quasi-
period of the profile variations $P\approx4^{h}$ determined above.

We conclude that the observed profile variations are consistent with the 
hypothesis that they are associated with photospheric NRPs of the primary, 
Aa$^1$, of the $\delta\,$Ori~A  system in the sector mode 
$(l,\mid m \mid)=(2,2)$. 
Since the profile features in the wavelet spectra move 
from the violet to the red part of the profile, $m <0$, 
yielding $(l,m)=(2,-2)$ for the NRP mode~\cite{TS97II}. 

Apart from these features of the dynamical wavelet spectra, which we suppose 
to be related to NRPs, these spectra also display weaker features in the 
violet and red line wings beyond the $\pm V\sin i$ band. 
These cannot be due to variations in the photosphere of Aa$^1$. Two 
explanations for these specific features of the line profile variations are 
possible. First, they may be related to processes in the stellar wind in 
the expanding atmosphere of Aa$^1$. Since (i) we have 
estimated the contributions from Aa$^2$ and Ab to the 
main line profiles to be relatively small, (ii) CIIII$\,\lambda\,$5696 
emission appears only in the spectra of the brightest 
hot supergiants (in our case, Aa$^1$), and (iii) the variability 
of the CIII emission is likely due to variability of photospheric absorption 
lines, and also exceeds the limits of the $\pm V\sin i$ band, we suggest that 
the variability of the main lines is primarily due to NRPs 
in the atmosphere of the Aa1 supergiant, and also to 
a variable contribution from the emission of the wind 
from the primary, modulated by its NRPs. Second, 
the variability beyond the $\pm V\sin i$ band could be 
related to the components Aa$^2$ and Ab. 

Two clearly pronounced parallel structures are visible in the dynamical 
wavelet spectrum of the CIII$\,\lambda\,$5696 differential line profiles 
(Fig.\ref{fig.dynWL50}), associated with the movement of broad features in 
the wavelet spectrum towards negative velocities. The right component moves 
with a velocity of $\approx 180 \, \mbox{km/s}$ to 
$\approx 40 \, \mbox{km/s}$ and is located fully (taking into account 
its own line-profile widths) within the $\pm V\sin i$ band. The band crossing 
time for this component is $\approx 6^h$, which substantially exceeds the 
crossing time for the components of the HeII$\,\lambda\,$4686 and 
HeI$\,\lambda\,$4713 lines related to NRPs. These structures are undoubtedly 
associated with some imhomogeneities in the atmosphere that contribute to the 
emission at the CIII$\,\lambda\,$5696 line frequencies, rather than directly with 
NRPs of the primary. The possibility of generating 
quasi-regular structures in a stellar wind originating 
from photospheric NRPs is indicated in~\cite{OC-puls}. 

In support of this idea, we note that the patterns  of the dynamical wavelet 
spectra related to gaseous imhomogeneities detected in the CIII emission, and 
also the wavelet spectrum features seen in other lines 
beyond the $\pm V \sin i$ band, are mutually complementary. 
Bringing into coincidence the wavelet spectra for the HeI$\lambda$4713 
photospheric line and the CIII$\,\lambda\,$5696 envelope line 
(Fig.~\ref{fig.dynWL50}) 
indicates an avoidance of the wavelet spectra; i.e. the absence of any 
features in the CIII$\lambda$5696 emission wavelet spectrum in the velocity 
range containing the features in the HeI$\lambda$4713 photospheric 
absorption wavelet spectrum. 

\begin{figure*}[!ht]
\centering
\includegraphics[height=10.0cm,width=14.0cm,angle=00]{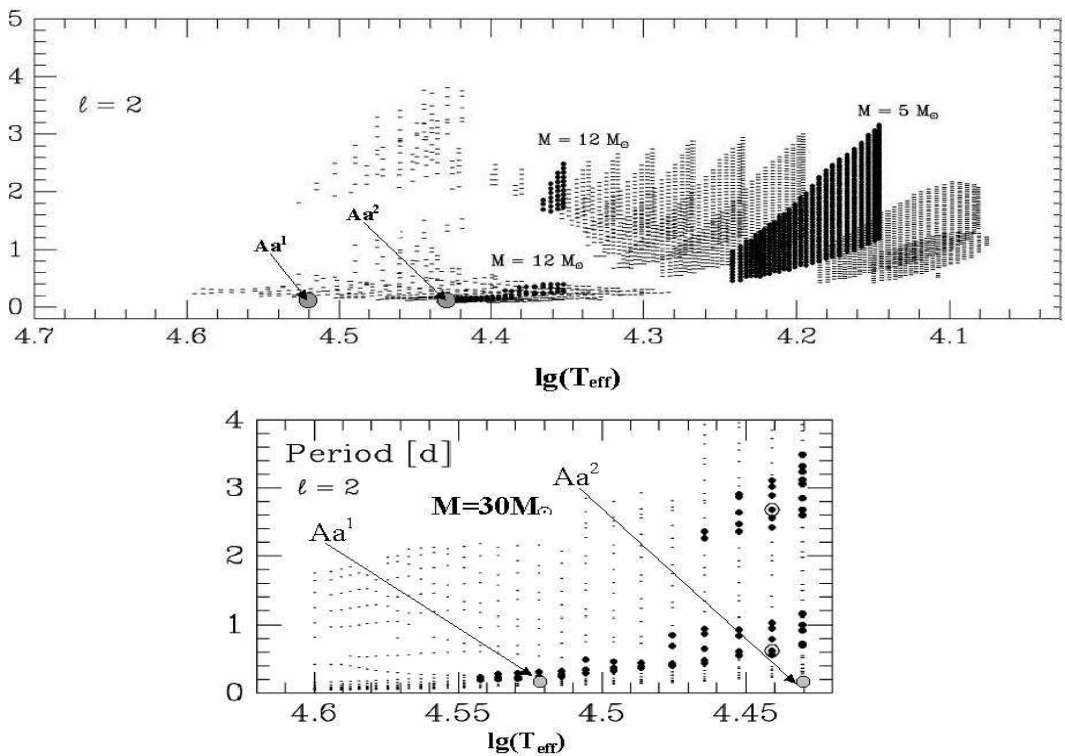}
\caption{\footnotesize Position of components Aa$^1$ and Aa$^2$ (grey circles) 
          in a diagram of $\log T_{\mbox{\scriptsize eff}}$ vs. pulsation 
          period $P$~\cite{Pam99}. The top plot shows the periods for 
          the pulsation modes $l=2$ for $\beta\,$Cep type stars and slowly 
          pulsating B stars; the small dots indicate the calculated periods, 
          and the bold dots the pulsation periods of main-sequence stars with 
          masses of 5$M_{\odot}$ and 12$M_{\odot}$. The bottom plot shows the 
          periods for pulsation modes $l=2$ for a main-sequence star with a 
          mass of 30$M_{\odot}$. 
         }
\label{fig.Teff-P}
\end{figure*}

To clarify the extent to which the obtained quasi-period for the line-profile 
variations in the spectrum of $\delta\,$Ori~A corresponds to possible 
pulsation periods of OB stars, we have plotted the positions of Aa$^1$ 
and Aa$^1$ on an diagram of effective temperature vs. pulsation period in the 
quadrupole mode $l=2$ for hot main-sequence stars 
(~\cite{Pam99}, Figs. 6 and 7]. The positions of the components in 
Fig.~\ref{fig.Teff-P} (top) are below the short-period boundary of the 
pulsation-instability strip for these stars, and do not correspond to the 
masses of the stars determined in\cite{Harvin02} (see Table). 
At the same time, the position of Aa1 in the diagram 
calculated for a high-mass main-sequence star with 
M =30 M. (Fig.~\ref{fig.Teff-P}, bottom) accurately corresponds 
to the pulsation-instability strip for this star. 

We suggest that, even in the case of substantial mass-loss from a star, its 
inner structure does not vary appreciably, and the periods of its NRPs 
correspond to pulsations for a main-sequence star whose mass is equal to the 
initial mass of the star up to the mass-loss stage. 

Confirmation of the reality of the obtained period and clarification of the 
nature of the line-profile variations requires spectra of $\delta\,$Ori~A 
obtained over two to three nights, making it possible to encompass four to 
six NRP cycles.

\section{CONCLUSION}
\label{s.Concl}

The following conclusions may be drawn from our 
observations. 

1. All the studied lines display variable profiles, 
with the variability amplitude being 0.5\%-1\%. This 
conclusion has validity level of 0.999. 

2. Large-scale components in the velocity interval 25-50 km/s are detected in 
   the dynamical wavelet spectrum of the HeII$\,\lambda$4686, 
   HeI$\,\lambda$4713,H., and CIII$\,\lambda$5696 line profile variations, 
   which move in the $-V\sin i \div V\sin i$ band for the primary of the 
   system, Aa$^1$, with a band crossing time of 4-5$^h$. 
   Some variable features that go beyond the band are 
   most likely due to emission imhomogeneities in the 
   wind from $\delta\,$Ori~Aa$^1$. The less bright components of 
   the system Aa$^2$ or Ab may also contribute to the 
   variability beyond the $-V\sin i \div V\sin i$ band. 
3. We have found short-period variations of the studied line profiles with 
   the characteristic times $\approx4{^h}.1$.. We present evidence 
   suggesting that these variations are associated with non-radial 
   pulsations of the O9.5II primary of the system, Aa$^1$, in the sector 
   mode $(l,m)=(2,-2)$.

{\footnotesize\it
\hspace*{-0.3cm}  
%
\section{ACKNOWLEDGEMENTS} 
The authors are grateful to V.E. Panchuk for assistance with the observations 
and to A.B. Schneiweiss for the Fourier spectra calculations. This work 
was supported by the Russian Foundation for Basic Research (project code 
05-02-16995a), a Presidential Grant of the Russian Federation in Support 
of Leading Scientific Schools (NSh-088.2003.3), and a Presidential Grant of 
the Russian Federation in Support of Young Candidates of Science (MC-
874.2004.2). One of the authors (G.G.V.) acknowledges the Korea MOST Foundation
(grant M1-0222-00-0005), as well as the KOFST and KASI 
programs (Brain Pool Program, Korea). 
}


\end{document}